\documentclass[letterpaper, 10 pt, conference]{ieeeconf}

\IEEEoverridecommandlockouts                              

\usepackage{comment}
\usepackage{cite}
\usepackage{amsmath,amssymb,amsfonts}
\usepackage{algorithmic}
\usepackage{graphicx}
\usepackage[export]{adjustbox}
\usepackage{booktabs}
\usepackage{textcomp}
\usepackage{xcolor}
\def\BibTeX{{\rm B\kern-.05em{\sc i\kern-.025em b}\kern-.08em
    T\kern-.1667em\lower.7ex\hbox{E}\kern-.125emX}}
\usepackage{url}
\usepackage[hidelinks]{hyperref}

\newcommand{\mcD}{\mathcal{D}}
\newcommand{\mcE}{\mathcal{E}}
\newcommand{\mcG}{\mathcal{G}}
\newcommand{\mcL}{\mathcal{L}}
\newcommand{\mcN}{\mathcal{N}}

\newcommand{\bc}{\mathbf{c}}

\newcommand{\bp}{\mathbf{p}}
\newcommand{\bq}{\mathbf{q}}
\newcommand{\bu}{\mathbf{u}}
\newcommand{\bx}{\mathbf{x}}
\newcommand{\bz}{\mathbf{z}}

\newcommand{\bA}{\mathbf{A}}
\newcommand{\bB}{\mathbf{B}}
\newcommand{\bC}{\mathbf{C}}
\newcommand{\bD}{\mathbf{D}}
\newcommand{\bE}{\mathbf{E}}
\newcommand{\bI}{\mathbf{I}}
\newcommand{\bK}{\mathbf{K}}
\newcommand{\bM}{\mathbf{M}}
\newcommand{\bQ}{\mathbf{Q}}
\newcommand{\bR}{\mathbf{R}}
\newcommand{\bS}{\mathbf{S}}
\newcommand{\bT}{\mathbf{T}}

\newcommand{\bdelta}{\boldsymbol{\delta}}
\newcommand{\btheta}{\boldsymbol{\theta}}
\newcommand{\bxi}{\boldsymbol{\xi}}
\newcommand{\blambda}{\boldsymbol{\lambda}}

\newcommand{\bpi}{\boldsymbol{\pi}}
\newcommand{\bomega}{\boldsymbol{\omega}}

\newcommand{\bPhi}{\boldsymbol{\Phi}}
\newcommand{\bGamma}{\boldsymbol{\Gamma}}
\newcommand{\bLambda}{\boldsymbol{\Lambda}}
\newcommand{\bSigma}{\boldsymbol{\Sigma}}

\newcommand{\bell}{\boldsymbol{\ell}}

\newcommand{\1}{\boldsymbol{1}}
\newcommand{\0}{\boldsymbol{0}}

\newcommand{\real}{\ensuremath{\mathbb{R}}}
\newcommand{\rf}{\text{ref}}
\newcommand{\blk}{\text{blkdiag}}
\newcommand{\diag}{\text{diag}}

\title{\LARGE \bf Modeling and Stability of Power Systems With Coupled Data Center and Grid Dynamics}

\author{Guido Cavraro
\thanks{G. Cavraro is with the Power Systems Engineering Center, National Laboratory of the Rockies. Email: {\tt guido.cavraro@nlr.gov}}
\thanks{The work of E. Dall'Anese was supported in part by the National Science Foundation Award 2504084. This work was authored in part by the National Laboratory of the Rockies for the U.S. Department of Energy (DOE), operated under Contract No. DE-AC36-08GO28308. This work was supported in part by the Laboratory Directed Research and Development (LDRD) Program at the National Laboratory of the Rockies. The views expressed in the article do not necessarily represent the views of the DOE or the U.S. Government. The U.S. Government retains and the publisher, by accepting the article for publication, acknowledges that the U.S. Government retains a nonexclusive, paid-up, irrevocable, worldwide license to publish or reproduce the published form of this work, or allow others to do so, for U.S. Government purposes.}
}

\begin{document}

\maketitle

\thispagestyle{empty}
\pagestyle{empty}

\begin{abstract}
The growing deployment of data centers poses challenges for power-system operation. However, data centers possess flexible resources that can be leveraged to mitigate their impact on the grid and provide grid services. This paper develops a linear-affine dynamic model of a transmission network with multiple data centers, explicitly coupling electromechanical grid dynamics with data center power demand and computational backlog. Grid-aware computing policies are incorporated to represent how data centers adjust their computational power in response to grid frequency deviations and workload backlog. The proposed framework provides a tractable means of investigating the interactions between multiple data centers and the transmission grid and assessing the potential of flexible computing to support power-system stability.
\end{abstract}

\section{Introduction}

The rapid deployment of large data centers (DCs), driven largely by AI training and inference, is creating new challenges for power-system planning, operation, and stability~\cite{kercci2024impact}, e.g., rapid and correlated variations in DC power demand can affect grid frequency~\cite{shenoy2026stabilitypowersystemsai}.
To facilitate DC integration, several DC power-demand models~\cite{ko2026,vercellino2026measurement}, detailed digital twins of DCs~\cite{Athavale}, and electromagnetic transient (EMT) models~\cite{Brett} have been proposed. However, these models typically do not explicitly couple computational workload dynamics with grid-level quantities, thereby overlooking a key characteristic that distinguishes DCs from conventional large loads. Indeed, a DC's power consumption is shaped by its computational workload, including job-arrival rates, scheduling policies, and the computational throughput achievable per unit of energy~\cite{colangelo2026characterizing}. These computational processes exhibit their own dynamics, which interact with those of the power system.

Several research efforts are exploring how DCs can become \emph{grid-aware} and \emph{grid-interactive} assets that not only mitigate their impact on the grid but also actively support power-system operation~\cite{Gupta,Ko2026Mitigation, Ren2026, ZHANG2025124697}. DCs can leverage their inherent flexibility, including uninterruptible power supply (UPS) battery systems, cooling systems, and deferrable computational workloads, to provide grid services across different time scales~\cite{colangelo2026ai}. 
At fast time scales, frequency support can be provided by rapidly discharging UPS batteries or temporarily adjusting computational power~\cite{Tao2025}. 
At slower time scales, energy management can be achieved through workload scheduling, e.g., by deferring computationally intensive training jobs to periods of greater energy availability or more favorable grid conditions~\cite{colangelo2026ai} or energy prices~\cite{Xinyang_DC2026}. By coordinating these mechanisms, DCs could contribute to both short-term frequency regulation and longer-term energy management while accounting for the computational performance requirements of their workloads.

This work develops a framework for investigating the role of multiple DCs as flexible resources for supporting power-system stability. The remainder of the paper is organized as follows. Section~\ref{sec:system_model} develops a coupled dynamic model of a transmission network with multiple DCs. The model combines lossless linearized power-flow equations, generator swing equations, first-order turbine-governor dynamics, and fluid-queue models of DC computational backlogs~\cite{Natarajan2026}. Section~\ref{sec:control} introduces a grid-aware computing policy in which computational power is modulated as a function of computational backlog and grid frequency. The resulting closed-loop system is linear-affine and explicitly captures the interactions among network frequency dynamics, DC power demand, and computational workload evolution, enabling a range of analyses and design tasks. Numerical simulations on the IEEE 14-bus system are presented in Section~\ref{sec:sim}. Eigenvalue analysis assesses the small-signal stability of the coupled system, while time-domain simulations quantify its response to disturbances under different grid-aware computing policies. Finally, Section~\ref{sec:conclusion} presents the conclusions.

\paragraph*{Notation and Preliminaries}
Lowercase (uppercase) boldface letters denote column vectors (matrices). Sets are represented by calligraphic symbols. The symbol $\1$ denotes the vector of all ones, whereas $\bI$ denotes the identity matrix. The symbol $\0$ denotes a vector or matrix whose entries are all zero. To avoid ambiguity, the dimensions of $\1$, $\0$, and $\bI$ are explicitly indicated by subscripts when needed. The $k$-th entry of a vector $\bx$ is denoted by $x_k$. The matrix $\diag(\{x_n\})$ is a diagonal matrix whose diagonal entries are the elements of the sequence of scalars $\{x_n\}$, whereas $\blk(\cdot)$ denotes a block-diagonal matrix formed by concatenating the matrices listed within the parentheses.

\section{System Modeling}
\label{sec:system_model}

We consider a connected power transmission grid with $N$ buses buses collected in the set $\mcN = \{1, ..., N\}$. Power transmission lines are collected in the set $\mcE = {(m, n)}$, with $\mcE \subseteq \mcN \times \mcN$. 
The grid hosts $N_G$ generator buses collected in the set $\mcG$, $N_D$ DC campuses collected in the set $\mcD$, and $N_L$ loads collected in the set $\mcL$, with $N_G + N_D + N_L = N$.
Consider bus $n$; its deviation from the nominal voltage frequency $\omega_0$ is denoted as $\omega_n$, whereas its voltage angle is $\theta_n$. 
Let the vectors $\bomega_G, \bomega_D, \bomega_L$ ($\btheta_G, \btheta_D, \btheta_L $) collect the frequency deviations (voltage angles ) for generator buses, DC buses, and loads, respectively.
The nodal power injection are collected in the vectors  $\bpi_G, \bpi_D, \bpi_L$, with $\pi_n > 0 ( < 0 )$ if power is actually injected (absorbed) at bus $n$. Heed that $\bomega_G, \btheta_G, \bpi_G \in \real^{N_G}$, $\bomega_D, \btheta_D, \bpi_D \in \real^{N_D}$, $\bomega_L, \btheta_L, \bpi_L \in \real^{N_L}$. 

The bus susceptance matrix is denoted by $\tilde \bB$; $\tilde \bB$ is block partitioned as
$$\tilde \bB=
\begin{bmatrix}
\tilde \bB_{GG}&\tilde \bB_{GD}&\tilde \bB_{GL}\\
\tilde \bB_{DG}&\tilde \bB_{DD}&\tilde \bB_{DL}\\
\tilde \bB_{LG}&\tilde \bB_{LD}&\tilde \bB_{LL}
\end{bmatrix}.
$$
Heed that $\tilde \bB \1= \0$. In this work, power flow equations are approximated by the DC equation
\begin{equation}
\begin{bmatrix}
\bpi_G^\top & \bpi_D^\top & \bpi_L^\top    
\end{bmatrix}^\top 
= \tilde \bB
\begin{bmatrix}
\btheta_G^\top & \btheta_D^\top & \btheta_L^\top    
\end{bmatrix}^\top
\label{eq:DC_appx}
\end{equation}
By modeling loads as constant power (PQ) buses, the system dimension can be reduced via Kron reduction. Assuming $\tilde \bB_{LL}$ invertible, 
we obtain the
\begin{equation}
\begin{bmatrix}
\bpi_G \\ \bpi_D    
\end{bmatrix} = 
\bB
\begin{bmatrix}
\btheta_G \\ \btheta_D    
\end{bmatrix} 
+
\begin{bmatrix}
\tilde \bc_G \\ \bc_D    
\end{bmatrix} 
\end{equation}
where the Kron reduced susceptance matrix $\bB$ is defined as
$$\bB = \begin{bmatrix}
\bB_{GG} & \bB_{GD} \\
\bB_{DG} & \bB_{DD}    
\end{bmatrix}
$$
with
$$
\begin{aligned}
& \bB_{GG} =\tilde \bB_{GG}-\tilde \bB_{GL} \tilde \bB_{LL}^{-1} \tilde \bB_{LG},\\
& \bB_{GD} = \tilde \bB_{GD} -  \tilde \bB_{GL} \tilde \bB_{LL}^{-1} \tilde \bB_{LD},\\
& \bB_{DG} = \tilde \bB_{DG}- \tilde \bB_{DL} \tilde \bB_{LL}^{-1} \tilde \bB_{LG},\\
& \bB_{DD} = \tilde \bB_{DD}- \tilde \bB_{DL} \tilde \bB_{LL}^{-1} \tilde \bB_{LD}\\
&\tilde \bc_G =- \tilde \bB_{GL} \tilde \bB_{LL}^{-1}\bpi_L, \quad
\bc_D=- \tilde \bB_{DL} \tilde \bB_{LL}^{-1}\bpi_L.
\end{aligned}
$$
Notice that 
\begin{equation}
    \bB \1 = \0
\label{eq:kron_red}
\end{equation}
The generator dynamics are represented by the swing equation coupled with a first-order turbine-governor model:
\begin{subequations}
\begin{align}
& \dot \btheta_G = \bomega_G \\
& \bM_G \dot \bomega_G = - \bD_G \bomega_G + \bp_G - 
\bB_{G}
\begin{bmatrix}
\btheta_{G} \\ \btheta_{D}
\end{bmatrix}
+ \bc_G \\
& \bT_G \dot \bp_G = - \bp_G - \bR_G^{-1} \bomega_G + \bp_G^\rf
\end{align}    
\label{eq:generators}
\end{subequations}    
where $\bM_G$, $\bD_G$, $\bT_G$, $\bR_G^{-1} \in \real^{N_G \times N_G}$ are diagonal matrices collecting the generator inertias, the damping coefficients, the turbine-governor time constants, and the droop control coefficients. Further, $ \bc_G = \tilde \bc_G - \bell_G$ with $\bell_G$  collecting the load demands at the generator buses, and 
$
\bB_G = 
\begin{bmatrix}
\bB_{GG} & \bB_{GD}
\end{bmatrix}
$.
Finally, $\bp_G,\bp_G^\rf \in \real^{N_G}$ are the vectors of the generators'  mechanical powers and power setpoints.

Each bus in \(\mcD\) represents a DC campus with a colocated synchronous generator. Thus, we assume that each DC campus is served by a dedicated synchronous generator.
The equations modeling the behavior of buses in $\mcD$ are 
\begin{subequations}
\begin{align}
& \dot \btheta_D = \bomega_D \label{eq:dc_delta}\\
& \bM_D \dot \bomega_D = - \bD_D \bomega_D + \bp_D - \bell_D - \bB_D  
\begin{bmatrix}
\btheta_{G} \\ \btheta_{D}
\end{bmatrix}
+ \bc_D \label{eq:dc_swing} \\
& \bT_D \dot \bp_D = - \bp_D - \bR_D^{-1} \bomega_D + \bp_D^\rf \label{eq:dc_gov}\\
& \bS \dot \bell_D = \bell_D^0 + \bxi - \bell_D \label{eq:pdc}\\
& \dot \bq = \blambda - \bK(\bell_D - \bell_D^0) \label{eq:q}
\end{align}    
\label{eq:DC}
\end{subequations}    
where $\bM_D$, $\bD_D$, $\bT_D$, $\bR_D^{-1} \in \real^{N_D \times N_D}$ are diagonal matrices collecting the generator inertias, the damping parameters, the turbine-governor time constants, the droop control coefficients and 
$
\bB_D = 
\begin{bmatrix}
\bB_{DG} & \bB_{DD}
\end{bmatrix}
$. 
The vectors $\bp_D, \bp_D^\rf, \bell_D, \bell_D^0, \bxi \in \real^{N_D}$ collect the generator mechanical powers, the generator power setpoints, DC facility overall power demand, and the DC facility fixed (or varying at slower time scales) power demand associated with non-computational functions or inflexible computational demand, and the reference power used for computation, respectively.
The vectors $\bq,\blambda \in \real^{N_D}$ represent the computational-work backlogs in the DCs, measured in floating-point operations (FLOP), and the job arrival rate, measured in FLOP/s. Considering the DC at bus $n$, $k_n$, measured in $\mathrm{FLOP}/(\mathrm{s}\cdot\mathrm{W})$, represent the effective conversion factor from computational power to computational throughput. 
Since $\bell_D^0$ is fixed, the instantaneous power dedicated to computation in the DC at bus $n$ is $(\ell_D)_n - (p_D^0)_n$.
Hence, $k_n [(\ell_D)_n - (p_D^0)_n]$ is the job clearing rate. All the $k_n$'s are collected in the diagonal matrix $\bK \in \real^{N_D \times N_D}$.

Equations~\eqref{eq:dc_delta}--\eqref{eq:dc_gov} describe the electromechanical dynamics of the synchronous generator and its interaction with the DC load. 
Equation~\eqref{eq:pdc} describes the dynamics of the DC power demand, with $\bS$ a diagonal matrix collecting the DC demand time constants. 
A similar first order model of DC power demand has been proposed already in~\cite{Milano2026}. 

Equation~\eqref{eq:q} describes a fluid approximation of the computational workload evolution and its dependence on the power allocated to computation. The aggregate computational backlog at DC $n$ is represented by the continuous state $q_n$ as motivated in~\cite{Natarajan2026}.

Using Little's law~\cite{little1961proof}, the average waiting time $\tau_n$ of a job at DC located at the bus $n$ can be estimated from $q_n$ as
\begin{equation}
\tau_n = q_n/\lambda_n.
\label{eq:Little}
\end{equation}

Define the $N_G + N_D$ dimensional vectors
$$
\btheta = 
\begin{bmatrix}
\btheta_G\\
\btheta_D
\end{bmatrix}, 
\bomega = 
\begin{bmatrix}
\bomega_G\\
\bomega_D
\end{bmatrix},
\bp = 
\begin{bmatrix}
\bp_G\\
\bp_D
\end{bmatrix}, 
\bp^\rf = 
\begin{bmatrix}
\bp_G^\rf\\
\bp_D^\rf
\end{bmatrix}
$$
and the matrices 
\begin{align*}
& \bD = \blk(\bD_G, \bD_D), \;\bM = \blk(\bM_G, \bM_D) \\
& \bT = \blk(\bT_G, \bT_D), \; \bR^{-1} = \blk(\bR_G^{-1}, \bR_D^{-1}) \; 
\end{align*}
To resolve the ambiguity associated with the absolute voltage angles, essentially arising from~\eqref{eq:kron_red}, we adopt the center of inertia (COI) transformation. Define the COI angle $\theta_{\text{COI}} = (\1^\top \bM \btheta)/(\1^\top \bM \1)
$
and the angle difference vector $\bdelta = \btheta - \1  \theta_{\text{COI}}$.
Let $\bQ$ be a matrix whose columns span the linear space $
\mathcal S = \{\bdelta: \1^\top \bM \bdelta = \0\}
$
and let $\bz \in \real^{N_G + N_D -1}$ such that
\begin{equation}
\bdelta = \bQ \bz.
\label{eq:COI_transf}
\end{equation}
It can be shown that the dynamic of $\bz$ evolves according to
\begin{equation}
\dot \bz = \bPhi \bomega    
\label{eq:z}
\end{equation}
with $\bPhi = (\bQ^\top \bM \bQ)^{-1} \bQ^\top \bM$.


\section{Closed loop linear model.}
\label{sec:control}

By adapting their power demand to both grid conditions and computational workload requirements, data centers (DCs) can become grid-interactive assets that support power-system operation. To this end, we adapt the computational-power reference of the DC at bus $n$ according to
\begin{equation}
\xi_n(\omega_n,q_n) = \gamma_n q_n + \sigma_n \omega_n
\label{eq:control}
\end{equation}
where $\gamma_n$ and $\sigma_n$ are positive control gains.
The first term in \eqref{eq:control} introduces computational backpressure into the control law. An increase in workload backlog $q_n$ raises the computational-power reference, thereby increasing the computational service rate as the actual power demand adjusts. This feedback mechanism couples workload dynamics to electrical demand and helps mitigate sustained backlog growth when sufficient computational capacity is available.

The second term implements frequency dependent droop control. When the grid frequency falls below its nominal value, the controller reduces the computational power reference, lowering the electrical load and helping alleviate under frequency conditions. Conversely, when frequency rises above its nominal value, the reference increases, allowing the DC to consume more power and provide additional demand flexibility when generation is relatively abundant.

By defining  the  $N_D \times N_D$ matrices
$
\bGamma = \diag(\{\gamma_n\}), \bSigma = \diag(\{\sigma_n\}),
$
and the $N_D \times (N_G + N_D)$ matrix 
$\bLambda = 
\begin{bmatrix}
\0_{N_D \times N_G} &    \bSigma 
\end{bmatrix},
$
we can compactly write the reference demand for all DCs as
\begin{equation}
\bxi = \bGamma \bq + \bLambda \bomega.
\label{eq:comp_back}
\end{equation}
Combining~\eqref{eq:generators}, \eqref{eq:DC}, \eqref{eq:COI_transf}, \eqref{eq:z} with the control rule~\eqref{eq:comp_back} yields the closed loop linear affine dynamical system
\begin{equation}
\bC \dot \bx = \bA \bx + \bu  
\label{eq:dyn_sys}
\end{equation}
where the state vector is 
$$
\bx = 
\begin{bmatrix}
\bz^\top & \bomega^\top & \bp^\top & \bell_D^\top & \bq^\top   
\end{bmatrix}^\top.
$$
Define $\bC = \blk(\bI_{N_G + N_D - 1}, \bM, \bT, \bS, \bI_{N_D})$, 
$\bc = 
\begin{bmatrix}
    \bc_G^\top & \bc_D^\top
\end{bmatrix}^\top$, 
$\bE = 
\begin{bmatrix}
\0_{N_G \times N_D} & \bI_{N_D} 
\end{bmatrix}^\top
$, and 
$$
\bA = 
\begin{bmatrix}
\0 & \bPhi & \0 & \0 & \0 \\
-\bB \bQ & - \bD & \bI & - \bE & \0\\
\0 & -\bR^{-1} & -\bI & \0 & \0 \\
\0 & \bLambda & \0 & -\bI & \bGamma \\
\0 & \0 & \0 & -\bK & \0
\end{bmatrix}, 
\bu = 
\begin{bmatrix}
\0 \\
\bc \\
\bp^\rf \\
\bell_D^0 \\
\blambda + \bK \bell_D^0 
\end{bmatrix}.
$$
The model in \eqref{eq:dyn_sys} supports several complementary analyses. Time-domain simulations can characterize the response to grid contingencies under grid-aware computing policies. Eigenvalue analysis can be used to assess local stability and determine how the control gains affect the system's dynamic modes. The model also supports systematic controller design, for example, by optimizing the frequency-response and backlog-feedback gains to balance grid support against computational service quality.

The equilibrium DC demand comprises the baseline facility demand and the computational power required to process the incoming workload
$
\bell_D^* = \bell_D^0+\bK^{-1}\blambda.
$
If the equilibrium frequency deviations are zero\footnote{The equilibrium frequency deviation satisfies \(\bomega^*=\eta\1\) for some \(\eta\in\mathbb R\), as implied by the COI equilibrium condition. In the following, we assume that a secondary frequency controller regulates the system frequency to its nominal value, so that \(\eta=0\).}, then $\bxi^* = \bGamma\bq^*$, and consequently
$
\bq^*=\bGamma^{-1}\bK^{-1}\blambda.
$
Thus, the equilibrium backlog and, through \eqref{eq:Little}, the corresponding estimate of the job average waiting time depend on $\bGamma$ but not on $\bLambda$. In particular, increasing $\gamma_n$ reduces the equilibrium backlog at DC $n$. However, both $\bGamma$ and $\bLambda$ can affect the stability properties of the closed-loop system~\cite{Natarajan2026}.



\section{Numerical Results}
\label{sec:sim}

In this section, we use the linear system~\eqref{eq:dyn_sys} to assess the benefits of adopting the grid aware computing policy~\eqref{eq:comp_back} via small signal stability analysis and via contingency analysis after a loss of generation. A modified version of the IEEE 14 bus test feeder is used as a model of the power transmission grid of interest. The network considered hosts three synchronous generators $\mcG = \{1, 2, 6\}$, and two DC campuses with colocated generation $\mcD = \{3,8\}$. The remaining buses host loads. Load values and generator sizes are provided in the test bed data spreadsheet.  Powers are expressed in per unit ($S_{\text{base}} = 100$~MVA) and the grid nominal frequency is  $\omega_s = 2\pi \cdot 60$~rad/s. The generator buses and the DC buses parameters are reported in Table~\ref{tab:gen_params} and Table~\ref{tab:dc_params}, respectively

\begin{figure}[tb]
\centering
\includegraphics[width=0.7\columnwidth]{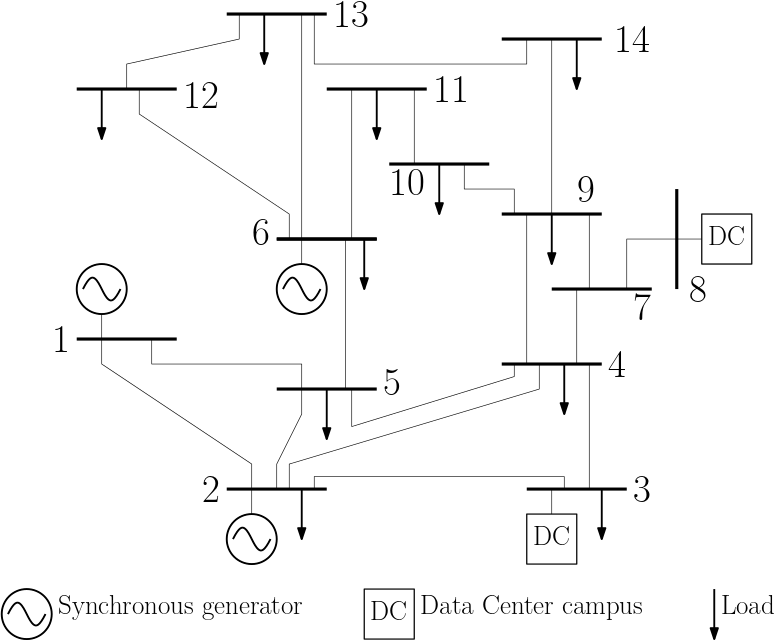}
\caption{Schematic representation of the IEEE 14 bus transmission test feeder.}
\label{fig:IEEE14}
\end{figure}

\begin{table}[t]
\centering
\caption{Generator parameters.}
\label{tab:gen_params}
\begin{tabular}{c l c c c }
\toprule
Bus & $M_n$ & $D_n$ & $T_n$ & $R_n^{-1}$ \\
    &  [pu$\cdot$s$^2$/rad] & [pu$\cdot$s/rad] & [s] & [pu$\cdot$s/rad] \\
\midrule
1   & 0.0265 &  0.00531 & 6.0 & 0.0531 \\
2   & 0.0212 &  0.00398 & 5.0 & 0.0531 \\
6   & 0.0186 &  0.00398 & 2.0 & 0.0531 \\
\midrule
3   & 0.00796 & 0.00265 & 1.0 & 0 \\
8   & 0.00531 &  0.00265 & 0.5 & 0 \\
\bottomrule
\end{tabular}
\end{table}

\begin{table}[t]
\centering
\caption{DC parameters.}
\label{tab:dc_params}
\begin{tabular}{c c c c c c}
\toprule
Bus & $S_n$ & $\lambda_n$ & $\kappa_n$ & $(\ell_D^0)_{n}$ & $\tau_{n}$ \\
    & [s] & [PFLOP/s] & [PFLOP/(s$\cdot$pu)] & [pu] & [s] \\
\midrule
3 & 0.5 & $5.94\times10^{4}$ & $1.414\times10^{5}$ & 0.18 & 5 \\
8 & 1.0 & $2.97\times10^{4}$ & $1.414\times10^{5}$ & 0.09 & 5 \\
\bottomrule
\end{tabular}
\end{table}

\subsubsection{Small signal stability analysis.}

\begin{figure*}[tb]
\centering
\includegraphics[width=0.90\textwidth]{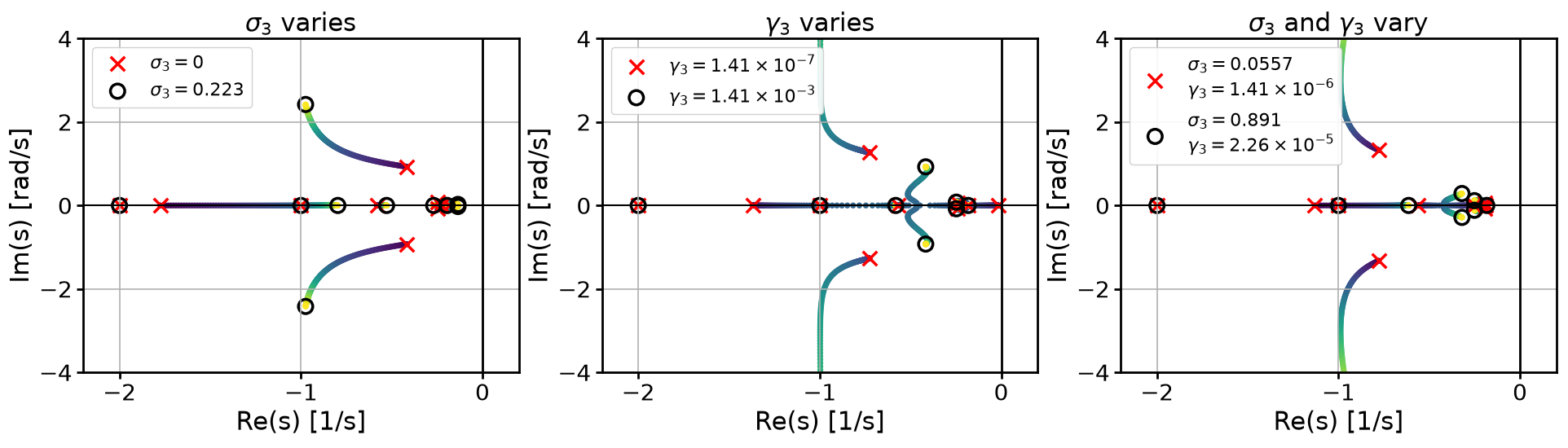}
\caption{Slow eigenvalues of $\bC^{-1} \bA$ as the computing-policy parameters of the DC at bus 3 are varied. Left: frequency-response gain $\sigma_3$ varied with the backlog gain $\gamma_3$ fixed. Center: $\gamma_3$ varied with $\sigma_3$ fixed. Right: both gains increased together. Markers indicate the eigenvalues at the start ($\times$) and end (o) of each sweep. Only the region near the imaginary axis is shown.
Increasing $\sigma_3$ alone moves a backlog-recovery mode toward the origin, while increasing both gains in proportion preserves the slow-mode stability margin.}
\label{fig:eig}
\end{figure*}

We use the developed model to carry out a small-signal stability analysis of the coupled grid–DC system. The equilibrium is asymptotically stable when all eigenvalues of $\bC^{-1} \bA$ have negative real parts. The eigenvalue locations also give the decay rate, oscillation frequency and damping of each mode. Figure~\ref{fig:eig} shows how the eigenvalues near the imaginary axis move as the two computing-policy parameters of the DC at bus 3 are varied: the frequency-response gain $\sigma_3$ and the backlog gain $\gamma_3$.
The left panel varies $\sigma_3$ from 0 to 0.22 pu/(rad/s) with $\gamma_3$ fixed. A stronger frequency response adds damping to the oscillatory frequency–governor modes. However, it also pushes a real eigenvalue, linked to recovery of the computation backlog, toward the origin. A DC that cuts its computing more aggressively during frequency events therefore takes longer to work through the resulting backlog, which shrinks the stability margin of the slow dynamics. The middle panel varies $\gamma_3$ with $\sigma_3$ fixed. 
For small $\gamma_3$, a slow real mode sits close to the origin, reflecting slow backlog recovery. As $\gamma_3$ grows, this mode moves left, then becomes an oscillatory pair whose frequency rises and whose damping decreases.
These observations suggest tuning both parameters together. Scaling both gains at the same time gave the results shown in the right panel and lets a DC offer stronger frequency support without slowing down the system's recovery. 

\subsubsection{Contingency Analysis}
\begin{figure*}[t]
\centering
\includegraphics[width=1\textwidth]{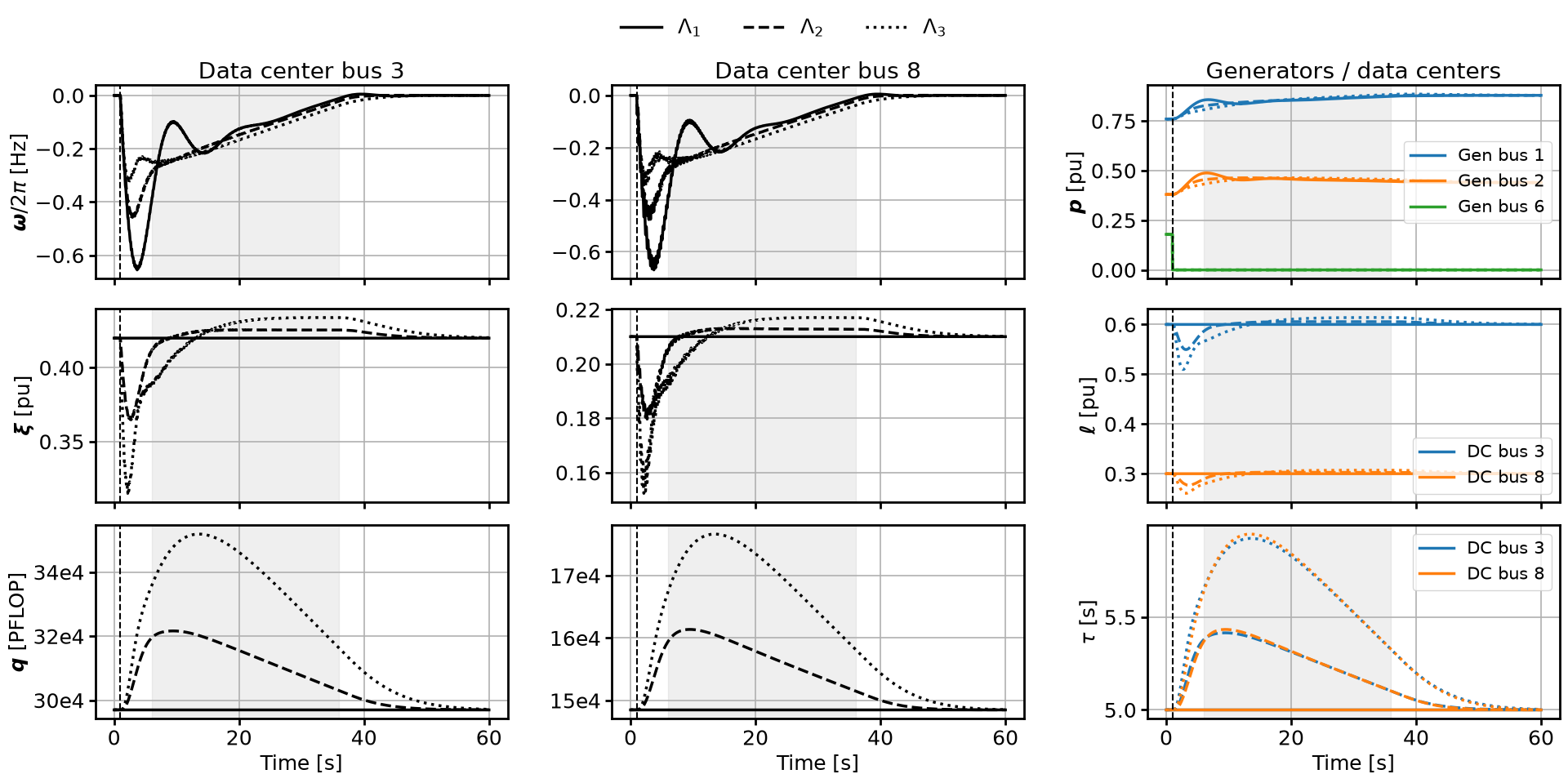}
\caption{System response  to the trip of the generator at bus 6 at $t = 1$ s. During the shaded interval (6–36 s), the remaining generators restore the power balance. Line styles denote the computing policies: grid-unaware $\bLambda_1$ (solid), and frequency-responsive $\bLambda_2$ (dashed) and $\bLambda_3$. Frequency-responsive computing raises the frequency nadir by up to about 50\% at the cost of temporary computing curtailment and modest increase in waiting time.}
\label{fig:step}
\end{figure*}

To assess how the grid-aware computing rule~\eqref{eq:comp_back} affects the system's response to large disturbances, we simulate a generator contingency. The results are reported in Figure~\ref{fig:step}.
The system starts at its pre-disturbance equilibrium, $\bx^* = -\bA^{-1} \bu$, so that any dynamics come only from the event.
At $t = 1$ s, the generator at bus is disconnected. From $t = 6$ s to $t = 36$ s, the remaining generators ramp their setpoints up, in proportion to their initial dispatch, until the lost power is replaced.
The run is repeated for three computing policies at both DCs: one grid-unaware policy with $\bLambda_1 = \0$, and two frequency-responsive policy with one ($\bLambda_3$) more aggressive than the other ($\bLambda_2$).
The results reveal a clear trade off between grid support and computational service quality. Under the grid unaware policy, the frequency nadir at the DC buses reaches approximately $-0.66$ Hz relative to nominal frequency, approaching levels at which under frequency load shedding may become a concern. The frequency-responsive policies reduce the magnitude of the frequency nadir by approximately 30\% and 50\%, respectively, by temporarily curtailing computational power.

This grid support comes at the cost of additional computational backlog. The estimated average job waiting time increases from its nominal value of 5 s to peaks of approximately 5.4 s and 5.9 s under the two frequency responsive policies. Thus, stronger frequency support results in longer waiting times. Within approximately 50 s of the disturbance, the remaining generators restore the power balance and the additional backlog is cleared.

\section{Conclusion}
\label{sec:conclusion}

This paper presents a linear-affine dynamic model of a transmission network with multiple DCs that couples power-system dynamics with DC power demand and computational backlog to capture the interactions between grid frequency, flexible computing demand, and workload evolution.
The model enables eigenvalue analysis to investigate the small-signal stability of the coupled system and time-domain simulations to evaluate its response to grid disturbances under different computing policies. These analyses provide a framework for assessing how grid-aware computing can contribute to frequency support and how control parameters influence the trade-off between grid response and computational backlog.
The proposed formulation provides a foundation for further investigating the role of multiple DCs as flexible resources in transmission networks. Future work will consider more detailed DC models, operational constraints, and systematic control-policy design to better quantify the potential of flexible computing to support reliable power-system operation.

\bibliographystyle{IEEEtran}
\bibliography{BibliographyGM}

\end{document}